\documentclass{revtex4} 
\usepackage{amsmath} 
\usepackage{amssymb} 
\usepackage{graphicx} 
\usepackage{color} 

\begin{document} 

\title{Hidden structures in quantum mechanics} 
\author{Vladimir Dzhunushaliev 
\footnote{Senior Associate of the Abdus Salam ICTP}} 
\email{vdzhunus@krsu.edu.kg} 
\affiliation{ASC, 
Department f{\" u}r Physik, Ludwig-Maximilians-Universit{\" a}t M{\" u}nchen, 
Theresienstr. 37, D-80333, Munich, Germany \\ 
and \\ 
Dept. Phys. and Microel. Engineer., Kyrgyz-Russian Slavic University, Bishkek, Kievskaya Str. 
44, 720021, Kyrgyz Republic} 

\date{\today}

\begin{abstract}
It is shown that some operators in quantum mechanics have hidden structures that are unobservable in principle. These structures are based on a supersymmetric decomposition of the momentum operator, and a nonassociative decomposition of the spin operator.
\end{abstract}


\maketitle

\section{Introduction} 

A majority of physicists these days believes that quantum mechanics does not have a hidden structure: Experiments have shown that a vast class of hidden variable theories is incompatible with observations. In essence, these theories assume existence of some hidden variables behind quantum mechanics, which could be measured in principle. In this paper we would like to show that in quantum mechanics there exist hidden structures based on a supersymmetric decomposition of the momentum operator, and a nonassociative decomposition of the spin operator. The constituents of this nonassociative decomposition are inaccessible to the experiment, because nonassociative parts of operators are unobservable.
\par 
In Ref. \cite{Gunaydin:1974} the attempt is made to introduce a nonassociative structure in quantum chromodynamics. As a consequence, not all states in the corresponding octonionic Hilbert space will be observable, because the propositional calculus of observable states as developed by Birkhoff and von Neumann \cite{birkhoff} can only have realizations as projective geometries corresponding to Hilbert spaces over associative composition algebras, whereas octonions are nonassociative. An observable subspace arises in the following way: Within Fock space there will be states that are observable (longitudinal, in the notation of Ref.~\cite{Gunaydin:1974}), which are the linear combinations of $u_0$ and $u^*_0$. Conversely, the states in transversal direction (spanned by $u_i$ and $u_i^*$) are unobservables ($u_0, u_0^*, u_i$ and $u_i^*$ are split octonions). 
\par 
A hidden structure in supersymmetric quantum mechanics is found in Ref. \cite{Dzhunushaliev:2007vg}. There, the Hamiltonian in supersymmetric quantum mechanics is decomposed as a bilinear combination of operators built from octonions, a nonassociative generalization of real numbers. 
\par 
In some sense, the Maxwell and Dirac equations have hidden nonassociative structures as well. In Ref's \cite{Gogberashvili:2005xb} and \cite{Gogberashvili:2005cp} it is shown that: (a) classical Maxwell equations can be written as the single continuity equation in the algebra of split octonions, and (b) the algebra of split octonions suffices to formulate a system of differential equations equivalent to the standard Dirac equation. 
\par 
In this paper we present hidden structures in traditional quantum mechanics. These structures are based on a supersymmetric decomposition of the momentum operator, and a nonassociative decomposition of the spin operator. We would like to emphasize that a "hidden nonassociative structure" presented here is not the same as a "hidden variable theory", because \textit{the nonassociative constituents of a hidden structure can not be measured in principle, in contrast to hidden variables which can be measured in principle. } 

\section{A supersymmetric decomposition of the momentum operator} 

The Poincar\'{e} algebra is defined with the generators $M^{\mu \nu}, P^\mu$ and the following commutator relations 
\begin{eqnarray} 
 \left[ P^\mu , P^\nu \right] &=& 0 , 
\label{1-10a}\\ 
 \left[ M^{\mu \nu} , P^\lambda \right] &=& 
 i \left( \eta^{\nu \lambda} P^\mu - \eta^{\mu \lambda} P^\nu \right) , 
\label{1-20a}\\ 
 \left[ M^{\mu \nu} , M^{\rho \sigma} \right] &=& 
 i \left( \eta^{\nu \rho} M^{\mu \sigma} + \eta^{\mu \sigma} M^{\nu \rho} - 
 \eta^{\mu \rho} M^{\nu \sigma} -\eta^{\nu \sigma} M^{\mu \rho} \right) 
\label{1-30a} 
\end{eqnarray} 
where $\eta_{\mu \nu} = \mathrm{diag} \left( +,-,-,- \right)$ is the Minkowski metric, $P^\mu$ are generators of the translation group, and $M^{\mu \nu} = x^\mu P^\nu - x^\nu P^\mu$ are generators of the Lorentz group. 
\par 
The simplest supersymmetric algebra is defined as follows 
\begin{eqnarray} 
 \left[ P^\mu , Q_a \right] &=& \sigma^\mu_{a \dot a} \bar Q^{\dot a} , 
\label{1-10}\\ 
 \left[ P^\mu , \bar Q^{\dot a} \right] &=& - \sigma^{\mu \dot a a} Q_a , 
\label{1-20}\\ 
 \left[ M^{\mu \nu} , Q_a \right] &=& - i \left( \sigma^{\mu \nu} \right)^{\;\;\dot b}_a \bar Q_{\dot b} , 
\label{1-30}\\ 
 \left[ M^{\mu \nu} , \bar Q^{\dot a} \right] &=& - i \left( \sigma^{\mu \nu} \right)^{\dot a}_{\;\;b} Q^{b} , 
\label{1-40}\\ 
 \left\{ Q_a , Q_{\dot a} \right\} &=& 2 \sigma^\mu_{a \dot a} P_\mu , 
\label{1-50}\\ 
 \left\{ Q_a , Q_b \right\} &=& \left\{ \bar Q_{\dot a} , \bar Q_{\dot b} \right\} = 0. 
\label{1-60} 
\end{eqnarray} 
where 
\begin{eqnarray} 
 \sigma^{\mu} &=& \frac{1}{4} \left( 
   \mathbf I_2, \vec \sigma 
 \right), 
\label{1-70}\\ 
 \bar \sigma^{\mu} &=& \frac{1}{4} \left( 
   \mathbf I_2, - \vec \sigma 
 \right) = \sigma_{\mu} 
\label{1-80} 
\end{eqnarray} 
having indices $\left( \sigma^\mu \right)_{a \dot a}$ and  $\left( \bar \sigma^\mu \right)^{a \dot a}$; 
$\vec \sigma = \left( \sigma^1, \sigma^2, \sigma^3 \right)$ are the Pauli matrices. The relation \eqref{1-50} can be inverted 
\begin{equation} 
 P_\mu = \frac{1}{4} \sigma^{a \dot a}_\mu \left\{ Q_a , Q_{\dot a} \right\}. 
\label{1-90} 
\end{equation} 
Equation \eqref{1-90} can be interpreted as a ``square root'' of the quantum mechanical momentum operator $P_\mu$. It allows us to bring forward the question: Is it possible to decompose other operators in quantum mechanics in a similar manner, for example, spin $\hat s_i$ and angular momentum $M^{\mu \nu}$? 
\par 
The undotted indices are raised with 
\begin{equation} 
 \epsilon^{ab} = i \sigma^2 = \left( 
         \begin{array}{cc} 
         0          &     -1     \\ 
         1          &     0      
         \end{array} 
    \right) . 
\label{1-100} 
\end{equation} 
The dotted indices are lowered with 
\begin{equation} 
 \epsilon_{\dot a \dot b} = - i \sigma^2 = \left( 
         \begin{array}{cc} 
         0          &     1     \\ 
         -1          &     0      
         \end{array} 
    \right) 
\label{1-110} 
\end{equation} 
and 
\begin{eqnarray} 
 \epsilon^{a b} &=& \epsilon^{\dot a \dot b} , 
\label{1-120}\\ 
 \epsilon_{a b} &=& \epsilon_{\dot a \dot b} . 
\label{1-130} 
\end{eqnarray} 

\section{A nonassociative decomposition of the spin operator} 

Let us consider split-octonion numbers, designated as $q_i$ ($i=1,2, \cdots 7$). Table \ref{oct} represents the chosen multiplication rules for the $q_i$. 
\begin{table}[h]
\begin{tabular}{|c|c|c|c|c|c|c|c|c|}                                                
\hline
&$q_1$ & $q_2$ & $q_3$ & $q_4$  & $q_5$ & $q_6$ & $q_7$         \\ 
\hline 
$q_1$& $-1  $ & $q_3$  & $-q_2$  & $-q_7$ & $q_6$  & $-q_5$ & $q_4$ \\ 
\hline
$q_2$ & $-q_3$ & $-1$   & $q_1$  & $-q_6$  & $-q_7$ & $q_4$  & $q_5$    \\ 
\hline
$q_3$ & $q_2$ & $-q_1$ & $-1$   & $q_5$  & $-q_4$  & $-q_7$ & $q_6$    \\ 
\hline
$q_4$ & $q_7 $ & $q_6$ & $-q_5$ & $ 1$   & $-q_3$  & $q_2$  & $q_1$    \\ 
\hline
$q_5$ & $-q_6$ & $q_7$  & $q_4$ & $q_3$ & $ 1$   & $-q_1$  & $q_2$    \\ 
\hline
$q_6$ & $q_5 $ & $-q_4$ & $q_7$  & $-q_2$ & $q_1$ & $ 1$   & $q_3$    \\ 
\hline
$q_7$ & $-q_4 $ & $-q_5$  & $-q_6$ & $-q_1$  & $-q_2$ & $-q_3$ & $ 1$    \\ 
\hline
\end{tabular}
\caption{The split-octonion multiplication table.} 
\label{oct}
\end{table}
\par
Split-octonions have the following commutators and associators 
\begin{eqnarray} 
 \left[ q_{i+3} , q_{j+3} \right] &=& - 2 \epsilon_{ijk} q_k , 
\label{2-10}\\ 
 \left[ q_i, q_j \right] &=& 2 \epsilon_{ijk} q_k  , 
\label{2-30}\\ 
 \left( q_{i+3}, q_{j+3}, q_{k+3} \right) &=& \left( q_{i+3} q_{j+3} \right) q_{k+3} - 
 q_{i+3} \left( q_{j+3} q_{k+3} \right) = 2 \epsilon_{ijk} q_7 
\label{2-40} 
\end{eqnarray} 
here $i,j,k = 1,2,3$. The commutator \eqref{2-30} shows that $q_i, i=1,2,3$ form a subalgabra. This subalgebra is called quaternion algebra $\mathbb H$; $q_i$ are quaternions. 
\par 
The commutator \eqref{2-30} can be rewritten in the form 
\begin{equation} 
 \left[ \frac{\imath}{2} q_i, \frac{\imath}{2} q_j \right] = 
 \epsilon_{ijk} \frac{\imath}{2} q_k , \quad \imath^2 = -1 
\label{2-50} 
\end{equation} 
which is similar to the commutator relationship for spin operators 
$\hat s_i = \frac{\sigma_i}{2}$  ($\sigma_i$ are Pauli matrices). It allows us to say that nonrelativistic spin operators have a hidden nonassociative structure \eqref{2-10}. The multiplication table \ref{oct} shows that the nonrelativistic spin operator can be decomposed as the product of two nonassociative numbers 
\begin{equation} 
 \frac{\imath}{2}q_i = - \frac{\imath}{4}\epsilon_{ijk} q_{j+3} q_{k+3} .
\label{2-60} 
\end{equation} 
In order to see it more concisely, let us represent the split-octonions via the Zorn vector matrices
\begin{equation}	
	\left(
	\begin{array}{cc}
		a				&	\vec x	\\
		\vec y	&	b
	\end{array}
	\right)
\label{2-70}
\end{equation}
where $a,b$ are real numbers and $\vec x, \vec y$ are 3-vectors, with the product defined as
\begin{equation}	
	\left(
	\begin{array}{cc}
		a				&	\vec x	\\
		\vec y	&	b
	\end{array}
	\right)
	\left(
	\begin{array}{cc}
		c				&	\vec u	\\
		\vec v	&	d
	\end{array}
	\right) =
	\left(
	\begin{array}{rl}
		ac + \vec x \cdot \vec v										&	\quad a \vec u + d \vec x -
		\vec y \times \vec v 																														 \\
		c \vec y + b \vec v + \vec x \times \vec u	&	\quad bd + \vec y \cdot \vec u
	\end{array}
	\right)
\label{2-80}
\end{equation}
Here, $( \cdot )$ and $[ \times ]$ denote the usual scalar and vector products.
\par
If the basis vectors of 3D Euclidean space are $\vec e_i, i=1,2,3$ with
$\vec e_i \times \vec e_j = \epsilon_{ijk} \vec e_k$ and $\vec e_i \cdot \vec e_j = \delta_{ij}$, then we can rewrite the split-octonions as matrices
\begin{eqnarray}	
	1 & = & \left(
	\begin{array}{ll}
		1				&	\vec 0	\\
		\vec 0	&	1
	\end{array}
	\right), \quad
	q_7 = - \left(
	\begin{array}{cc}
		1				&	\vec 0	\\
		\vec 0	&	-1
	\end{array}
	\right), \quad
\label{2-90}\\
	q_i & = & \left(
	\begin{array}{ll}
		0					&	- \vec e_i	\\
		\vec e_i	&	0
	\end{array}
	\right), 
\label{2-100}\\
	q_{i+3} &=& \left(
	\begin{array}{cc}
		0					&	\vec e_i	\\
		\vec e_i	&	0
	\end{array}
	\right)
\label{2-110}
\end{eqnarray}
here $i=1,2,3.$ Thus, the nonrelativistic spin operators have two representations: The first one is as Pauli matrices $\hat s_i = \frac{\sigma_i}{2}$ with the usual matrix product, and the second one is as Zorn matrices \eqref{2-100} $\hat s_i = \frac{\imath}{2} q_i$ with nonassociative product \eqref{2-80}. In the second case, the spin is decomposed into a product \eqref{2-60} of two unobservables $q_{j+3},q_{k+3}$.

\section{Quantum mechanical applications} 

In the supersymmetric approach, the operators $P^\mu$ with commutator relations \eqref{1-10a} are generators of the Poincar\'{e} group. We interpret these as quantum mechanical operators, and consider nonrelativistic quantum mechanics. Taking the decomposition \eqref{1-90} and \eqref{2-60} into account, we have 
\begin{eqnarray} 
 \hat P_i &=& \frac{1}{4} \sigma_i^{a \dot a} \left\lbrace 
   Q_a , Q_{\dot a} 
 \right\rbrace , 
\label{3-10}\\ 
 \hat s^i &=& - \frac{1}{4} \epsilon_{ijk} \left[ q_{j+3}, q_{k+3} \right] .
\label{3-30} 
\end{eqnarray} 
We offer the following interpretation of Eq's \eqref{3-10} and \eqref{3-30}: \textit{Quantum mechanics has hidden supersymmetric and nonassociative structures, which can be expressed through decomposition of classical momentum and spin operators, into bilinear combinations of some operators that are either supersymmetric or nonassociative.} 
\par 
Probably, such nonassociative hidden structure \textit{can not be found experimentally in principle}, because the nonassociative parts $q_{5,6,7}$ generate unobservables (for details of unobservability, see Ref. \cite{Dzhunushaliev:2007wu}). 

\section{Discussion and conclusions} 

We have shown that some quantum mechanical operators can be decomposed into supersymmetric and nonassociative constituents. The following questions outline further investigation in this direction: 
\begin{enumerate} 
	\item Does a 4D generalization of relations \eqref{3-10}-\eqref{3-30} exist? 
	\item Do $Q_a, Q_{\dot a}$ have dynamical equations ? 
	\item Is a similar nonassociative decomposition of quantum field theory possible? 
\end{enumerate} 
The first question can be formulated mathematically as follows: Find a nonassociative algebra $\mathcal R$, with commutators and associators 
\begin{eqnarray} 
 \left[ R^\mu , \tilde R^\nu \right] &=& 2 M^{\mu \nu}, 
\label{4-1}\\ 
 \left[ M^{\mu \nu}, M^{\rho \sigma}\right] &=& \imath \left( \eta^{\nu \rho} M^{\mu \sigma} + 
    \eta^{\mu \sigma} M^{\nu \rho} - 
 \eta^{\mu \rho} M^{\nu \sigma} -\eta^{\nu \sigma} M^{\mu \rho} \right)  , 
\label{4-3}\\ 
 \left( P^\mu, P^\nu, P^\rho \right) &=& 
    \left( P^\mu P^\nu \right) P^\rho - P^\mu \left( P^\nu P^\rho \right) = 
 2 \epsilon^{\mu \nu \rho \sigma} P_\sigma 
\label{4-4} 
\end{eqnarray} 
where $P^\mu = \text{either } R^\mu \text{ or } \tilde R^\mu$. Also, ask whether a linear representation for $R^\mu, \tilde R^\nu$ exists. This question arises, because supersymmetric operators $Q_a, Q_{\dot a}$ have a linear representation 
\begin{eqnarray} 
 i Q_a &=& \frac{\partial}{\partial \theta^a} - i \sigma^\mu_{a \dot a} \bar \theta^{\dot a} \partial_\mu, 
\label{4-10}\\ 
 i \bar Q_{\dot a} &=& 
    - \frac{\partial}{\partial \theta^{\dot a}} + i \theta^{a} \sigma^\mu_{a \dot a} \partial_\mu . 
\label{4-20} 
\end{eqnarray} 
These operators are generators of translation, in a superspace with coordinates 
$z^M~=~\left( x^\mu, \theta_a, \theta_{\dot a} \right)$, where $\theta_a, \theta_{\dot a}$ are Grassmanian numbers obeying 
\begin{equation} 
 \left\lbrace \theta_a, \theta_b \right\rbrace = 
    \left\lbrace \bar \theta_{\dot a}, \bar \theta_{\dot b} \right\rbrace  = 
    \left\lbrace \theta_a, \bar \theta_{\dot a} \right\rbrace = 0. 
\label{4-30} 
\end{equation} 
The second question from above is important, because for any classical quantum mechanical operator $L$ we can write the Hamilton equation 
\begin{equation} 
  \frac{d L}{dt} = \frac{\partial L}{\partial t} + i \left[ H, L \right] 
\label{4-50} 
\end{equation} 
where $H$ is a Hamiltonian. For nonassociative parts of operators, however, there is an obstacle for such equation: Because $H$ is generated from a product of two or more constituents, their nonassociativity demands to define the order of brackets in the product of $HL$ and $LH$. 
\par 
In Ref. \cite{okubo1995} the question is considered: What is the most general nonassociative algebra $\mathcal A$ which is compatible with Eq. \eqref{4-50}. Let us consider the consistency condition 
\begin{equation} 
  \frac{d (xy)}{dt} = \frac{d x}{dt} y + x \frac{d y}{dt} , 
\label{4-60} 
\end{equation} 
which requires validity of 
\begin{equation} 
  \left[ H, xy \right] = x \left[ H, y \right] + \left[ H, x \right] y. 
\label{4-70} 
\end{equation} 
The validity of Eq. \eqref{4-70} is not obvious for a general algebra. There exists the following theorem 
\par 
\underline{Theorem} (Myung \cite{myung}) 
\par 
To have a necessary and sufficient condition for 
\begin{equation} 
  \left[ z, xy \right] = x \left[ z, y \right] + \left[ z, x \right] y. 
\label{4-80} 
\end{equation} 
for any $x,y,z \in \mathcal A$, it is required that $\mathcal A$ is flexible and Lie-admissible, i.e. 
\begin{eqnarray} 
 (x,y,z) &=& - (z,y,x), 
\label{4-90}\\ 
 \left[ \left[ x,y \right] z \right] + \left[ \left[ z,x \right], y \right] + 
 \left[ \left[ y,z \right], x \right] &=& 0. 
\label{4-100} 
\end{eqnarray} 
\par 
Finally, a few notes about the third question. In Ref. \cite{Gunaydin:1974} the idea is offered that by quantization of strongly interacting fields (in particular in quantum chromodynamics), nonassociative properties of quantum field operators may arise. In \cite{Gunaydin:1974} it is proposed that a quark spinor field $\psi$ can be presented as a bilinear combination of usual spinor fields $\psi_i$ and nonassociative numbers (split octonions) $q_i$. Both ideas, in Ref. \cite{Gunaydin:1974} and here, are qualitatively similar: Quantum operators can be decomposed into nonassociative constituents. 
\par
An unsolved problem exists in quantum chromodynamics: confinement. The challenging property is that a quark-antiquark pair cannot be separated. Physically speaking, this means that a single quark cannot be observed. Mathematically, the problem is that we do not know the algebra yet, which models field operators for strongly interacting fields (gluons for quantum chromodynamics)\footnote{The gluons carry out strong interaction between quarks.}. One can suppose that the unobservability of quarks can be connected with a non-associative structure of the algebra of gluon operators. 

\section{Acknowledgments} 
I am grateful to the Alexander von Humboldt Foundation for financial support, to V. Mukhanov 
for invitation to Universit\"at M\"unchen for research, and to J. K\"oplinger for fruitful discussion. 

\appendix 

\section{Some definitions for nonassociative algebras} 
\label{na} 

Textbooks for nonassociative algebras are e.g.~Ref's \cite{schafer}, physical applications of nonassociative algebras in physics can be found in Ref's.~\cite{okubo1995} and \cite{baez}. 

A nonassociative algebra $\mathcal A$ over a field $K$ is a $K$-vector space $\mathcal A$ equipped with a $K$-bilinear map $\mathcal A \times \mathcal A \rightarrow \mathcal A$. 

An algebra is unitary if it has a unit or identity element $I$ with $Ix = x = xI$ for all $x$ in the algebra. 

An algebra is power associative if $x^n$ is well-defined for all $x$ in the algebra and all positive integer $n$. 

An algebra is alternative if $(xx)y = x(xy)$ and $y(xx) = (yx)x$ for all $x$ and $y$. 

A Jordan algebra is commutative and satisfies the Jordan property $(xy)(xx) = x(y(xx))$ for all $x$ and $y$. 

The associator is defined as follows 
\begin{equation} 
  \left( x,y,z \right) \equiv \left( x y \right) z - x \left( y z \right) . 
\label{a1-10} 
\end{equation} 
Any algebra obeying the flexible law 
\begin{equation} 
  \left( x,y,z \right) = - \left( z,y,x \right) 
\label{a1-40} 
\end{equation} 
is called a flexible algebra. 
\par 
Any algebra obeying the Jacobi identity 
\begin{equation} 
 \left[ \left[ x,y \right] , z \right] + 
    \left[ \left[ z,x \right] , y \right] + 
    \left[ \left[ y,z \right] , x \right] = 0 
\label{a1-50} 
\end{equation} 
is called a Lie-admissible algebra.

\end{document}